\documentclass[sigconf,nonacm]{acmart} 
\AtBeginDocument{%
  }

\begin{document}
\title{Collaboration Signatures: Tracking Long-Horizon Dynamics of Cross-Meeting Collaboration Networks}

\author{Shengqi Zhu}
\authornote{Work completed during internship at Microsoft.}
\email{sz595@cornell.edu}
\orcid{0009-0002-5660-3241}
\affiliation{%
  \institution{Cornell University}
  \city{Ithaca}
  \state{New York}
  \country{USA}
}

\author{Sheshera Mysore}
\email{smysore@iesl.cs.umass.edu}
\orcid{0000-0001-6453-3242}
\affiliation{%
  \institution{Microsoft Office of Applied Research}
  \city{Redmond}
  \state{Washington}
  \country{USA}
}

\author{Bahareh Sarrafzadeh}
\email{basarraf@microsoft.com}
\orcid{0000-0001-8765-870X}
\affiliation{%
  \institution{Microsoft Office of Applied Research}
  \city{Redmond}
  \state{Washington}
  \country{USA}
}

\begin{abstract}
Enterprise platforms have received increasing attention as a core component of social networks and the Web.
However, existing work mostly relies on fragmented logs of small units of work, like emails and chats, and focuses on who connects with whom without sufficient account of collaboration content and dynamics.
This work systematically explores tracking collaborations across sequences of meetings as a new approach to recovering and interpreting their long-horizon dynamics, describing in detail how joint efforts move forward from one meeting to the next.
Linking 45,426 meetings in a large multinational technology company based on full transcripts and 1.6 million extracted agenda items, we establish a \textit{cross-meeting collaboration network} of 6,065 collaborations. The network recovers the natural shapes of workplace collaborations: meetings, as snapshots of ongoing collaboration, link into largely \textit{tree-shaped} structures, while the work content forms nested \textit{directed acyclic graphs (DAGs)}.
Further, we explore core dynamics of how these networks evolve over time, examining three established perspectives: task division, team composition, and temporal continuity.
Each explains part of the dynamics but does not capture the distinctness of long-horizon workplace collaboration: e.g., work flows in a consistent direction between a few pairs of subjects, yet a collaboration's subject explains little of how its meetings are conducted.
Instead, accumulating the agenda, interaction, and language of meetings along each recovered tree yields a \textit{signature} of the collaboration, a distinctive profile that picks out the collaboration of a held-out meeting at 4.5 times chance.
This signature models collaboration better than team or task views, individually or combined, in next-meeting prediction and tree-level fitting.
We discuss the implications of these cross-meeting dynamics for designing office applications and support tools, and call for next-generation Web systems that represent collaborations over their long horizon.
\end{abstract}


\maketitle

\section{Introduction}
\label{sec:introduction}

Much of today's collaborative work unfolds on enterprise platforms, through recurring online meetings, chats, and shared documents~\cite{teevan2021new,defilippis2020collaborating}.
These have become a core part of the Web and of people's social networks, where connections among employees are close, continuous, and organized around shared work.
As collaboration proceeds, pieces of work are raised, split, discussed, revisited, handed on, and eventually concluded (or not).
Meeting platforms now record these meetings at scale for summarization, analytics, and AI assistance~\cite{zhong2021qmsum,hosseinkashi2024meeting,lewis2020retrieval}, and research on AI at work is turning toward collective productivity, calling for systems that represent a group's shared goals and context~\cite{butler2025microsoft}.
Together, these records open a window onto how collaborative work develops over weeks and months, where multiple meetings as ``snapshots'' of the collaboration together document the long-horizon groupwork.

However, while records are discrete and unlabeled, the collaborations behind them flow continuously and latently: related meetings may lie days apart, involve varying groups of people, and describe the same work in different stages.
More fundamentally, existing records rarely let us ask a key question about collaboration: does each collaboration have its own way of working?
Large-scale Web research on workplace collaboration has mostly modeled who connects with whom~\cite{yang2021effects,yu2023large,kossinets2006empirical}, while studies of how work is divided, staffed, and coordinated~\cite{cataldo2006identification,valentine2017flash,arrow1995membership} rarely follow a collaboration across sessions where work is discussed.
These views implicitly reduce a collaboration to its people, its task, or its latest meeting; a full picture therefore requires recovering how meetings are connected and how a specific collaboration sets up its own style.

In this work, we set out to (re)construct and interpret such long-horizon collaborations as a \emph{cross-meeting collaboration network}, focusing on three key questions:

\begin{itemize}
  \item[RQ1.] Can cross-meeting collaboration networks be recovered reliably from discrete, unlabeled meeting records, and what network structure do they form?
  \item[RQ2.] What explains how these networks change over time, and how far do established perspectives on collaboration go?
  \item[RQ3.] What (if anything) describes each collaboration's unique, recognizable dynamics (``signatures'')?
\end{itemize}

\begin{figure*}[t!]
  \centering
  \includegraphics[width=0.85\textwidth]{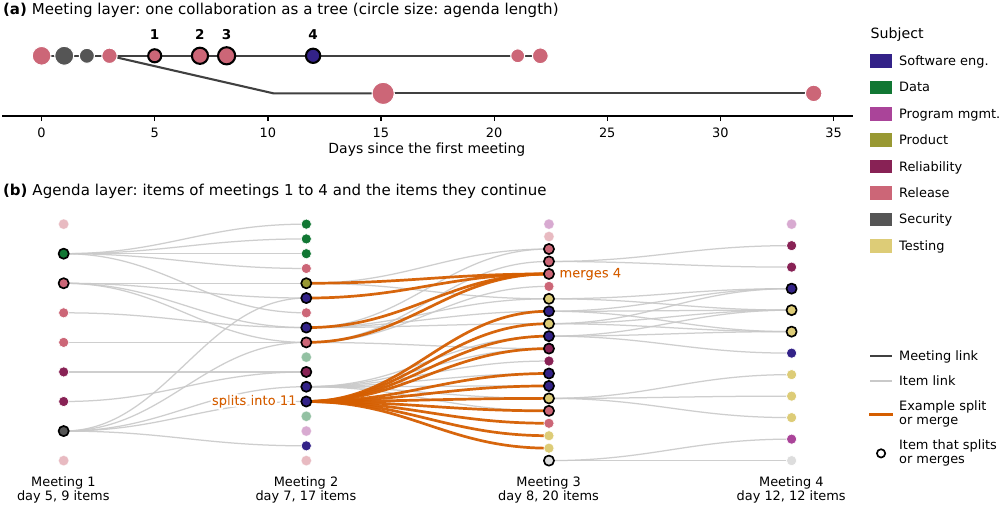}
  \caption{Both layers of one recovered collaboration. The meeting layer traces how the collaboration proceeds, and the agenda layer how its work is divided and recombined, as items split and merge between meetings. (a) Its twelve meetings, linked into a tree (circle color denotes main subject; circle size denotes agenda length; \S\ref{subsec:meeting-layer}). (b) Agenda items of the numbered meetings, colored by subject, and the links between them; one split and one merge are highlighted in orange, outlined circles denote items that split or merge, and pale circles denote items with no link to the meetings shown; meetings held within hours of each other are drawn slightly apart (\S\ref{subsec:agenda-layer}).}
  \Description{Two panels are shown. The top panel shows twelve meetings as circles along a time axis of 35 days, colored by main subject and joined into a tree with one branch; four consecutive meetings are numbered. The bottom panel shows the agenda items of these four meetings as columns of colored dots joined by thin gray links; highlighted links show one item of the second meeting splitting into eleven items of the third, and four items of the second meeting merging into one item of the third.}
  \label{fig:reconstruction}
\end{figure*}

To answer them, we link each meeting to the earlier meeting whose work it continues and each agenda item to the earlier items it continues, across 45,426 transcribed meetings of a large multinational technology company.
The result has two layers (Figure~\ref{fig:reconstruction}): a meeting layer that traces how a collaboration proceeds, and an agenda layer that traces how its work is divided and recombined.

First, the reconstruction recovers 6,065 multi-meeting collaborations whose two layers corroborate each other (99.0\% of meeting links are supported by separately judged item links).
The cross-meeting network takes a dual shape: meetings form trees, each typically continuing a single earlier meeting, while collaborations branch as they grow; the work they carry forms a dense directed acyclic graph, where agenda items split and merge across meetings.

On these trees, we examine three established perspectives on change, task division~\cite{malone1994interdisciplinary}, team composition~\cite{arrow1995membership}, and temporal continuity~\cite{konradt2021teams}, and find that each explains a real part of the dynamics, yet none accounts for what is distinctive about each collaboration.
Task division gives work a direction (e.g., Software Engineering $\rightarrow$ Release) and an ending, as work is recorded as completed mainly in the last meeting, but subjects enter in no staged order.
Team composition explains what persists: communication style, the way members talk and share the floor in meetings, varies far more between collaborations than between subjects, and newcomers bring the incoming subject with them.
Temporal continuity holds only in part: collaborations recur more than they chain, as earlier meetings reveal nearly as much about the next meeting as the latest one does.

Finally, we incorporate the three perspectives into a view of the collaboration itself, represented by its \emph{signature}, an identifiable profile based on how seemingly common per-meeting features accumulate along the tree.
A collaboration's signature identifies the collaboration of a held-out meeting at 4.5 times chance even among collaborations on the same subject.
About two thirds of it is brought in by participants, mainly as a style of talk, while the rest belongs to the collaboration, mainly in how it organizes its floor and its agenda; it is inherited by the branches a collaboration splits into, and emerges within its first meetings.
As a model of change, the signature anticipates the next meeting and fits whole trees better than the three perspectives, alone or combined.
A collaboration, on this view, carries a slowly drifting signature against which work is re-divided, flows between subjects, and comes to an end.
Hence, future assistants and analytics should be designed to draw on a collaboration's long-horizon record beyond the latest meeting.

Overall, our contributions include:

\begin{itemize}
  \item The cross-meeting collaboration network, recovered from discrete, unlabeled meeting records with two layers: meeting trees with sparse merges that carry a dense directed acyclic graph of agenda items (Section~\ref{sec:recovering}).
  \item An account of change in which task division, team composition, and temporal continuity each explain a measured part of the dynamics (Section~\ref{sec:change}).
  \item The collaboration signature, a unit of description made possible only by linked records: distinctive beyond subject and people and partly the collaboration's own, with implications for understanding collaboration and designing Web systems that support it (Sections~\ref{sec:signature} and~\ref{sec:discussion}).
\end{itemize}

\section{Related Work}
\label{sec:related-work}

\paragraph{Collaboration on enterprise platforms.}
Communication logs have long served to map networks inside organizations~\cite{kossinets2006empirical}.
The shift to remote and hybrid work made enterprise platforms the main site of collaboration~\cite{teevan2021new,lane2024teams} and opened it to measurement at scale: meetings became more frequent and shorter~\cite{defilippis2020collaborating}, collaboration networks more static and siloed~\cite{yang2021effects}, and new employees built their networks differently~\cite{yu2023large}.
Platform studies describe group chat~\cite{wang2022group}, collaborations across time zones \cite{mok2023challenging, sun2024rhythm}, multitasking in remote meetings~\cite{cao2021large}, and video interaction~\cite{bergmann2022meeting,brucks2022virtual}; generative AI assistants have also reduced time spent on email but not in meetings~\cite{dillon2025shifting}, calling for AI that supports collective work~\cite{butler2025microsoft}.
These studies take the person, tie, channel, or single meeting as their unit. In contrast, we follow collaboration and the work it carries across meetings.

\paragraph{Temporal and group dynamics.}
Temporal network research represents interactions as time-stamped links~\cite{holme2012temporal} and has followed groups that persist by changing their members~\cite{palla2007quantifying}.
Web and data-mining studies treat group interactions as units~\cite{benson2018simplicial} that recur with a bias toward recent ones~\cite{benson2018sequences}, much as people repeat earlier choices~\cite{anderson2014dynamics}, while first-order models misrepresent dynamics with memory~\cite{rosvall2014memory,scholtes2017network}.
Individuals keep social signatures in how they communicate, even as their contacts change~\cite{saramaki2014persistence}; we ask whether collaborations carry their memory and signatures at the tree level.

\paragraph{Recovering latent structure from Web records.}
Many Web records leave their structure implicit: search tasks that span sessions~\cite{kotov2011modeling,wang2013learning}, chat and email conversations~\cite{elsner2008you,yeh2006email,zhang2018making}, and news story lines that branch and merge~\cite{shahaf2012trains} have been recovered from their content.
Where a medium records structure, such as the reply trees of discussion threads~\cite{kumar2010dynamics,gomez2008statistical} or the declared branches of version control~\cite{bird2009promises}, the shape follows from it.
Meeting records offer little scaffolding beyond recurring meeting series, so the shape of the recovered network is an empirical question (Section~\ref{subsec:shape}), and language-model link judgments must be checked against independent signals~\cite{ziems2024can}.

\paragraph{How collaborative work changes.}
Research on collaborative work explains change through tasks, composition, and time.
Task-based accounts treat work as interdependent activities whose dependencies must be coordinated~\cite{malone1994interdisciplinary,wildman2012task}; studies of software development derive these needs from traces of shared work and find that they shift as work proceeds~\cite{cataldo2006identification}, and flash organizations restructure roles along the way~\cite{valentine2017flash}.
Composition-based accounts follow how groups gain and lose members and how newcomers bring new knowledge~\cite{arrow1995membership,choi2005old}; temporal accounts follow how coordination develops over successive phases~\cite{konradt2021teams} and how routines recur while changing~\cite{feldman2003reconceptualizing}; and multilevel theory asks which collective properties are composed from members and which emerge from their interaction~\cite{kozlowski2000multilevel}.
These accounts have rarely been tested on the same records; we compare them on the same collaborations.

\paragraph{Meetings as records of collaboration.}
Meeting science classifies the purposes of meetings and how teams facilitate them~\cite{allen2014understanding,lehmann2013sequential}, and computational work has built corpora for summarization~\cite{carletta2005ami,janin2003icsi,zhong2021qmsum}, related meeting features to participants' ratings~\cite{zhou2021role,hosseinkashi2024meeting}, and examined how tools represent meeting goals~\cite{scott2024mental}, in each case one meeting at a time.
Because simple persistence baselines often rival complex models on the Web~\cite{cheng2014can,poursafaei2022towards}, and prediction is most informative as a test of explanation~\cite{hofman2017prediction}, we test descriptions of collaboration by held-out prediction against persistence and permutation baselines.

\section{Cross-Meeting Collaboration Networks}
\label{sec:recovering}

\subsection{Source data: collection of meetings}
\label{subsec:meetings-agendas}

Our source data, collected under employee consent and institutional approval, consist of 45,426 transcribed online meetings held over five full months, from August to December 2025, in a large multinational technology company, with 32,209 distinct participants; the median meeting lasts 29.3 minutes and has four participants.\footnote{We describe how the records were collected, protected, and analyzed in Appendix~\ref{app:data}; no name or record text appears in this paper, and all results are aggregates.}
Following a collaboration through these records requires recovering (1) which earlier meeting each meeting continues and (2) which meetings thereby form one collaboration.

To describe what each meeting works on, a language model (GPT-5.5, on a privacy-compliant internal deployment) reads its transcript and extracts a structured agenda of atomic items, 1,626,259 in total and 35.8 per meeting on average.
Each item records what was discussed, to whom it is attributed, its temporal stage relative to that of the meeting (before, during, or planned after it), its status (completed, ongoing, or unknown), and one of fourteen subject fields\footnote{We annotate the subject field of each agenda item from a fixed set of fields of work, e.g., \textit{Software Engineering} or \textit{Business and Finance}; Appendix~\ref{app:data} lists all fourteen.}, and each meeting also receives a main subject (Appendix~\ref{app:data}).
Through this key design, agendas capture \textit{what a collaboration works on}, separate from \textit{how its meetings take place}, leading to the two-layer network constructed next.

\begin{figure*}[t]
  \centering
  \includegraphics[width=0.8\textwidth]{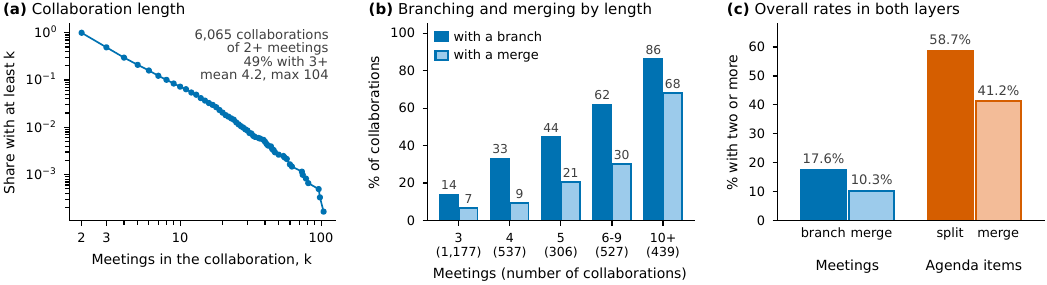}
  \caption{Shape of the two layers. Most collaborations are short and proceed as one thread, branching as they grow, while agenda items they carry split and merge. (a) Share of the 6,065 collaborations (two or more meetings) with at least k meetings (\S\ref{subsec:meeting-layer}). (b) Share of collaborations with a branch (dark) and with a merge (light), by number of meetings (collaborations in parentheses; \S\ref{subsec:meeting-layer} and Appendix~\ref{app:reconstruction}). (c) Share of meetings and items with a successor that have two or more (branch or split), and of those with a predecessor that have two or more (merge); meeting links are tree links and genuine merges (\S\ref{subsec:agenda-layer}).}
  \Description{Three panels are shown. Panel (a) is a log-log plot of the share of collaborations with at least k meetings, falling steadily from one to below one in a thousand at about one hundred meetings. Panel (b) is a bar chart in which the share of collaborations with a branch rises from 14 percent at three meetings to 86 percent at ten or more, and the share with a merge, lower in every length class, rises from 7 to 68 percent. Panel (c) is a bar chart: 17.6 percent of meetings with a successor branch and 10.3 percent of meetings with a predecessor merge, against 58.7 percent of agenda items that split and 41.2 percent that merge.}
  \label{fig:structure}
\end{figure*}

\subsection{Two-layer reconstruction}
\label{subsec:reconstruction}

With the meetings and their agendas, we represent collaboration as a \emph{cross-meeting collaboration network}, or cross-meeting network for short, with two layers.
In the meeting layer, a directed link $(m', m)$ states that meeting $m$ directly continues the work of an earlier meeting $m'$; and in the agenda layer, a link $(i', i)$ states that item $i$ continues an earlier item $i'$.
Item links are sought only between linked meetings, so the agenda layer is nested in the meeting layer. Both layers are acyclic, since links point forward in time.

\paragraph{Generating candidates.}
To infer latent collaboration continuation, we combine \textit{who attends} with \textit{what is discussed}.
An earlier meeting is a candidate predecessor if and only if it took place within the preceding time window (21 days) and shares a sufficiently similar participant pool (overlap count $\geq2$, and roster Jaccard similarity $\geq\frac{1}{3}$). A GPT-5.5 model then reads both meetings' info, compares the agendas, and predicts on a Likert scale how likely the later meeting directly continues the work of the earlier one (Appendix~\ref{app:data}).

\paragraph{From candidates to trees.}
We attach each meeting to its most recent candidate rated 4 or higher.
Most meetings are found to have a single firm predecessor: counting every candidate that qualifies, 77.5\% of the meetings that continue earlier work have only one, and 83.9\% of the accepted links receive the highest rating.
This reflects the tree shape of meeting sequences, which we examine specifically in Section~\ref{subsec:shape}: linking each meeting to its predecessor makes each collaboration, a connected set of at least two linked meetings, a tree, and the rarer meetings with a second predecessor that brings work of its own are recorded as merges.
Meetings that join no collaboration, one-off or unlinked, are excluded from the analysis.
Within each linked pair of meetings, dense and lexical similarity then propose the earlier items that each item may continue, and the language model keeps the continuations it judges with high confidence (Appendix~\ref{app:reconstruction}).
Figure~\ref{fig:reconstruction} illustrates both layers for one recovered collaboration.

\paragraph{Validating the links.}
While there is no ground truth for grouping, two further checks support the links.
A manual review by the authors of 100 sampled candidate pairs agreed with the linker's rating and its stated rationale for 98, and found the other two ambiguous.
The two layers also corroborate each other: 99.0\% of the meeting links between meetings with agendas are supported by at least one high-confidence item link.

\subsection{The shape of collaboration}
\label{subsec:shape}

Our first major finding is that the cross-meeting network takes a dual shape: meetings link into trees with sparse merges, while the work they carry forms a dense directed acyclic graph.
The trees give each collaboration a clear membership, order, and ancestry, on which the analyses of Sections~\ref{sec:change} and~\ref{sec:signature} build, while the DAG records how its work is divided and recombined beyond individual meetings.

\subsubsection{Meetings form trees with sparse merges}
\label{subsec:meeting-layer}
First, meetings typically have a strong parent-child relation, and most collaborations proceed as one linear thread (80.6\% never branch): 82.4\% of the meetings that are continued have a single successor, and 89.7\% of those continuing earlier work have a single predecessor (Figure~\ref{fig:structure}c).
Many of the 6,065 collaborations are short, but around half (46.4\%) have depth $\geq3$, counting the meetings along the longest path. The mean total meetings is 4.2, and the longest contains 104 meetings (Figure~\ref{fig:structure}a), spaced 4.9 days apart at the median.
As collaborations grow, they branch: 39.5\% of those with three or more meetings contain a meeting continued by two or more, and 86.1\% of those with ten or more (Figure~\ref{fig:structure}b).
The occasional but fundamental branching, universal across depths, is the defining feature of the tree model.

In addition to key branching, rare \textit{merges} complement the trees.
A meeting may also have another notable predecessor candidate, which we record as a genuine merge when each of the two parents accounts for $>2$ of the meeting's items that the other does not (Appendix~\ref{app:reconstruction}).
As merges are rather sparse, the meeting layer keeps the properties of a tree: membership in a collaboration, depth and height, etc. For instance, depth along the longest path through secondary parents would change for only 1.7\% of meetings, and thus the additional parent is recorded with no qualitative perturbation.

Overall, the meeting layer is a forest of trees, complemented by sparse genuine merges and local bridges.

\subsubsection{Agenda items form a dense directed acyclic graph}
\label{subsec:agenda-layer}
Second, the work carried along these trees is divided far more finely: where meetings branch, items split.
Of the items carried into a later meeting, 58.7\% continue into more than one item, and of the items that continue earlier work, 41.2\% draw on more than one, against 17.6\% and 10.3\% for meetings (Figure~\ref{fig:structure}c).
The splits are local (Figure~\ref{fig:reconstruction}b): for 95.8\% of the items with several successors, all successors lie in the same next meeting, and when a collaboration branches, only 16.7\% of the items passed on are taken up by more than one branch.
This division is not an artifact of how finely agendas are itemized: the share of items that split varies little with agenda length, from 54\% in the shortest fifth of agendas to 60\% in the longest. A single piece of work therefore cannot be traced as one item from meeting to meeting; what persists is the collaboration that carries it.

\subsection{Overview of the cross-meeting network}
\label{subsec:two-layers}

Altogether, 19,376 meeting links join 25,441 non-singleton meetings into 6,065 collaborations, and 537,087 item links run between the agendas of linked meetings.
The two layers complement each other in design, and also show distinct forms: \textit{a forest of tree-shaped meeting sequences carries a dense directed acyclic graph of work items}, re-divided along each link.
This heterogeneity is intriguing, especially as existing networks have not given emphasis to a model that combines meeting arrangements and work content, which we show to follow different dynamics. Person-tie networks of \textit{who interacts with whom}~\cite{kossinets2006empirical,yang2021effects}, for instance, collapse each meeting into pairwise ties between participants, losing both \textit{which meetings continue which} and \textit{what work they carry}.
In the vocabulary of higher-order networks~\cite{benson2018simplicial}, each meeting is a group interaction, the meeting layer orders these interactions into trees, and the agenda layer is a flow of work through them.
Change can thus occur in who meets and how they interact, and in what work is taken up and how it is divided.
In the next section, we examine these dynamics within the network, starting from three well-established views of workplace collaboration, which locate the source of change in the task, the team, or the continuity of work over time, and asking how far each accounts for how collaborations change.

\section{What delineates collaboration over time?}
\label{sec:change}
\label{subsec:views-design}

As we identify each collaboration from discrete meeting data, we ask: what drives the collaboration forward? What are the fundamental real-world dynamics that shape continued projects?
Along the recovered trees, the agenda takes up new subjects and re-divides its work from one meeting to the next, the collaborating team (roster) gains and loses participants, etc.
Research on collaborative work offers three established perspectives on such change: it follows the division and progression of \textit{tasks}~\cite{malone1994interdisciplinary,cataldo2006identification}, the composition of the \textit{team}~\cite{arrow1995membership,choi2005old}, or the \textit{continuity/trajectory} of work over time~\cite{konradt2021teams,feldman2003reconceptualizing}.

\paragraph{Testing established perspectives.}
Each perspective locates the source of what comes next in one view of a meeting:\footnote{We describe the features and models in detail in Appendix~\ref{app:prediction-details}.}
the task view describes the agenda through the number of items, their focus, and the shares at each stage and status;
the team view describes who attends and how they talk through transcript features, such as turn-taking, overlapping speech, the concentration of speaking time, and linguistic features (e.g., hedging, politeness, and first-person plural ``we'');
and the trajectory view places the meeting in its collaboration, through its depth and the intervals between meetings, which only the reconstructed network provides.
If one perspective captured the dynamics, its expectations should hold in the records,
and its view should anticipate the next meeting beyond simply carrying the current setup forward.
We therefore test each perspective against both its own expectation and the next meeting; where a perspective proves incomplete, it points to what else shapes a collaboration's course.

\paragraph{Selecting comparable measures.}
To compare the perspectives on common ground, we narrow the features down to a small, interpretable set of measures of what a meeting takes up and how it is conducted.
We establish 10 total measures, each posed as whether the next meeting lies above the corpus median: 4 of the agenda (its length in items, the share of items completed, the share planned for after the meeting, and its focus, the mean similarity between its items) and 6 of communication (turn length, number of speakers, overlapping talk, first-person plural framing, question rate, and unequal airtime; Appendix~\ref{app:team-details}).
Beyond the features of single meetings, we test what the order of meetings contributes by comparing each collaboration with random reorderings of its own meetings, which keep what it contains but remove the sequence.
Intervals resample whole collaborations, since meetings of the same collaboration resemble one another (Appendix~\ref{app:prediction-details}).
For each perspective, we state what it leads us to expect, then ask what it accounts for and what it leaves open.

\subsection{Task division: work flows downstream, yet subjects enter in no staged order}
\label{subsec:task-division}

A common picture in project management frameworks and the relevant tooling is that work moves through stages, as in the continuous delivery pipelines that carry software from build through testing to release~\cite{shahin2017continuous}.
Collaborations would then move through their subjects in a characteristic order, pass work downstream, and shift from planned toward completed work as they proceed.
Our results support only part of this intuitive picture.

\begin{figure}[t]
  \centering
  \includegraphics[width=0.8\columnwidth]{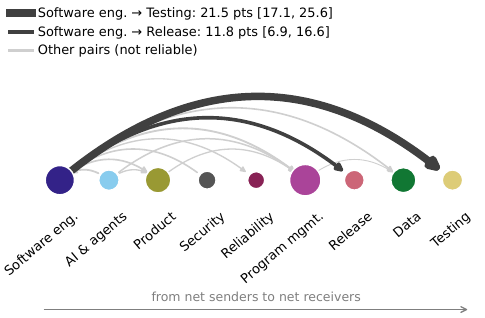}
  \caption{Net direction of item links for the twelve subject pairs most often joined by meeting links, from net senders (left) to net receivers (right). Most traffic between subjects runs both ways, and only two pairs show a significant net flow, from software engineering to testing and to release (\S\ref{subsec:task-division}). Node size represents cross-subject item links and arc width for net links. Dark arcs mark the two directions whose 95\% interval excludes zero.}
  \Description{An arc diagram with nine subjects as colored circles on a horizontal line, ordered from software engineering on the left to testing on the right. Thin gray arcs connect many pairs; two thick dark arcs run from software engineering to testing, with a net share of 21.5 percentage points, and from software engineering to release, with 11.8 points.}
  \label{fig:direction}
\end{figure}

First, instead of ``moving on'' from a distinct upstream to downstream, collaborations largely accumulate subjects over their course in vague ways.
Among the 393 collaborations that reach a ninth meeting, the distinct subjects discussed rise from 2.6 to 6.1 while each meeting covers about 2.6, and from the seventh meeting on, about nine in ten of a meeting's subjects were discussed earlier (Figure~\ref{fig:task-details}a).
Between some subjects, however, work does flow downstream.
Across the twelve subject pairs that are most often joined by meeting links, most item traffic runs both ways: no subject's sent and received links differ by more than 11\% of its total.
Two pairs stand out as significant (Figure~\ref{fig:direction}): of the item links between software engineering and testing, the share running toward testing exceeds the share running back by 21.5 percentage points (95\% CI 17.1 to 25.6), and between software engineering and release by 11.8 points (6.9 to 16.6).

\begin{figure*}[t!]
  \centering
  \includegraphics[width=0.8\textwidth]{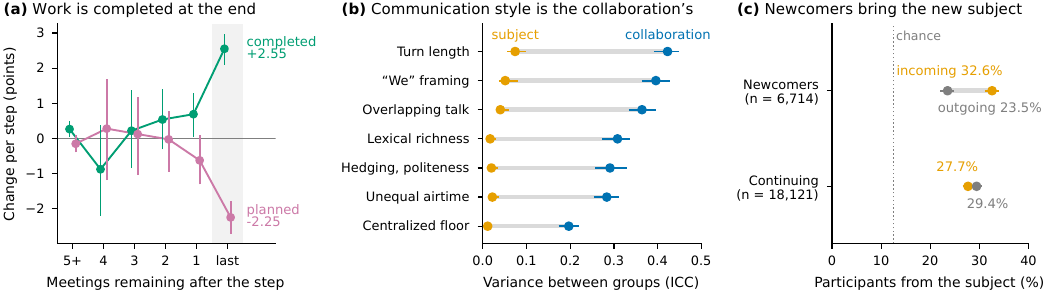}
  \caption{Task and team components. Work is recorded as completed mainly in the last meeting, communication style varies far more between collaborations than between subjects, and newcomers bring the incoming subject with them. (a) Change per step in the completed and planned shares of the agenda, by meetings remaining after the step (last: the step into the last meeting; \S\ref{subsec:task-division}). (b) Variance between collaborations and between main subjects (ICC) for the seven composite measures of communication style, defined in \S\ref{subsec:team-composition} and Appendix~\ref{app:team-details} (965 collaborations with at least six meetings). (c) At a change of main subject, share of participants whose earlier work lies mainly in the incoming or outgoing subject; dotted line: expected from subject frequencies (\S\ref{subsec:team-composition}). Intervals are 95\%, resampling collaborations.}
  \Description{Three panels. Panel a shows the change per step in the completed and planned shares by meetings remaining; both stay near zero until the last step, where the completed share rises by 2.55 points and the planned share falls by 2.25 points. Panel b is a dot plot in which, for each of seven measures of communication style, the share of variance between collaborations (0.20 to 0.42) is far larger than the share between subjects (0.01 to 0.07). Panel c is a dot plot: among newcomers, 32.6 percent come mainly from the incoming subject and 23.5 percent from the outgoing one; among continuing participants, the two shares are 27.7 and 29.4 percent; chance is 12.5 percent.}
  \label{fig:components}
\end{figure*}

Does this flow imply a staged order in which collaborations take up their subjects?
Compared with random reorderings of each collaboration's meetings, no subject's mean entry position moves by more than 0.046 on a scale from the first meeting (0) to the last (1) across 2,986 collaborations, and Software Engineering itself, like Security and Release, enters slightly later than random order would place it (Figure~\ref{fig:task-details}b).

Finally, the shift from planned toward completed work is concentrated at the end.
The completed share rises by 0.34 percentage points per step between consecutive meetings, but by 2.55 points in the step into a collaboration's last observed meeting, where the planned share falls by 2.25 (Figure~\ref{fig:components}a).
The shift is larger for endings well inside the observation window (Figure~\ref{fig:task-details}c), which suggests that collaborations record their work as done when they stop meeting, while throughout their course the average meeting records 63.9\% of its items as ongoing.

In summary, task division accounts partly for the direction of work and how collaborations end; however, subjects enter in no staged order and most pairs carry two-way traffic, suggesting that work is most often back and forth, much as coordination needs cross team boundaries and shift as the work proceeds~\cite{cataldo2006identification}.

\subsection{Team composition: participants persist, and newcomers bring new subjects}
\label{subsec:team-composition}

Accounts of membership dynamics reported that groups commonly gain and lose members as their tasks and circumstances change~\cite{arrow1995membership}.
If teams were formed around the task at hand, a collaboration that changes its subject would change its participants too, and its meetings would be conducted much like other meetings on the same subject.
The records point the other way on both counts, with one exception among those who join.

Communication style, the way a collaboration's members talk and share the floor in meetings, is much more a collaboration-specific attribute than a subject-specific one.
We summarize it with seven composite measures, each the mean of the standardized transcript features describing one construct (detailed in Appendix~\ref{app:team-details}).
Four describe the structure of interaction: turn length, how long participants hold the floor; overlapping talk, how often a turn begins before the previous one ends or is a brief acknowledgment; unequal airtime, how unevenly speaking time, turns, and words are divided among speakers; and a centralized floor, how far the handoffs from one speaker to the next pass through a single participant who acts as a hub.
Three describe language: lexical richness (varied and long words), hedging and politeness, and first-person plural framing (``we'' as against ``I'').
A case study of two collaborations illustrates the range (Appendix~\ref{app:team-details} and Figure~\ref{fig:style-cases}): one holds short, overlapping turns routed through a hub, while the other takes long turns with unequal airtime but no hub, in richer language framed as ``we''.
Across 12,050 meetings in 965 collaborations, differences between collaborations account for 20\% to 42\% of the variance of each of the seven composite measures of communication style, and between main subjects for at most 7\% (Figure~\ref{fig:components}b), consistent with prior team classifications that separate communication structure from task type~\cite{wildman2012task}.

Participants generally stay when the subject changes.
Although 43.4\% of meeting links join meetings with different main subjects, the share of a meeting's participants who also attended the previous one is nearly the same whether the subject changes (73.8\%) or not (76.2\%).
Those who do join, however, tend to bring the incoming subject.
Newcomers at a change of subject come mainly from the incoming subject in 32.6\% of cases and from the outgoing subject in 23.5\%, against 12.5\% expected from subject frequencies, whereas participants present at both meetings show no such tilt (27.7\% against 29.4\%; Figure~\ref{fig:components}c).
A collaboration thus changes its subject largely with the same people, and the people it adds come from where it is heading.

Team composition thus explains what persists through change, namely a collaboration's communication style and core participants, and how new subjects arrive, carried in by newcomers, echoing previous accounts of newcomers as sources of new knowledge~\cite{choi2005old}.

\subsection{Continuity along the trajectory: meetings recur more than they chain}
\label{subsec:temporal-continuity}

A third explanation looks to the time aspect itself as meetings proceed along the collaboration.
If meetings take up where the previous one left off, the latest meeting should carry most of what can be known about the next, as first-order models of network dynamics~\cite{rosvall2014memory,scholtes2017network} represent.
Successive agendas, however, overlap only partly in reality.
On average, 46.3\% of a later meeting's items continue an item from the predecessor meeting, and 27.5\% of the earlier meeting's items are taken up, both lower bounds since the item linker cannot find every continuation.
Neither share varies with the gap between meetings (Appendix~\ref{app:prediction-details}), suggesting that the overlap reflects how a collaboration organizes its agenda more than how much time has passed.

How much more does the latest meeting reveal than other meetings of the same collaboration?
At the 7,867 positions with at least three past meetings, the latest meeting's value indicates whether the next meeting lies above the corpus median with a mean AUC over the ten measures of 0.686, and the two meetings before it with 0.674 and 0.660.
A meeting drawn at random from the same collaboration, earlier or later, still gives 0.646, retaining about $\frac{4}{5}$ of the latest meeting's information above chance.
Time, then, contributes only a modest recency and a partial chaining of agendas, but most of what one meeting reveals about the next is carried by any meeting of the same collaboration.

Side by side, the three perspectives keep returning to \textit{the collaboration itself} while the work of each meeting is partly renewed.
If so, the collaboration should be described as a unit of its own, distinct from its single meetings, its participants, and its subjects.

\section{Signatures of Collaborations}
\label{sec:signature}

\begin{figure*}[t!]
  \centering
  \includegraphics[width=0.8\textwidth]{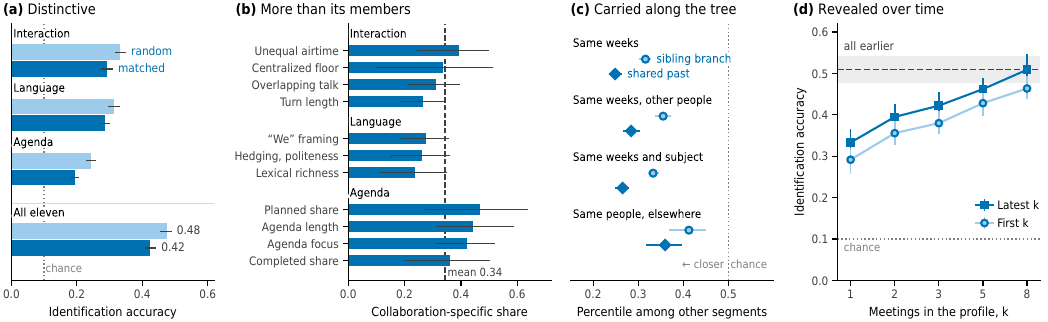}
  \caption{Signatures of collaborations. A collaboration's signature is distinctive beyond its subject and people, partly its own, inherited along its tree, and visible within its first meetings. (a) Share of 6,242 held-out meetings matched to their collaboration among ten candidates, by signals in the profile, against random and matched competitors (same subject, weeks, headcount, and duration; \S\ref{subsec:distinctive}). (b) Share of each measure's reliable profile not predicted by participants' profiles elsewhere or the subject, by kind of signal (956 collaborations; \S\ref{subsec:collaboration-specific}). (c) Mean percentile of a sibling branch, and of the shared past, among segments of other collaborations (1,969 branch points; 281 for the same people elsewhere; \S\ref{subsec:inherited}). (d) Identification of the last four meetings of 331 collaborations from the first or latest $k$ meetings, or all earlier ones (dashed; \S\ref{subsec:horizon}). Dotted lines mark chance.}
  \Description{Four panels. Panel a is a bar chart: interaction structure, language, and agenda alone identify a held-out meeting's collaboration about 19 to 33 percent of the time, and all eleven measures about 48 percent against random competitors and 42 percent against matched competitors, with chance at 10 percent. Panel b shows horizontal bars of the collaboration-specific share for eleven measures grouped into interaction, language, and agenda, from about 0.24 for lexical richness to about 0.47 for the planned share, with a mean near 0.34; all three language measures lie below the mean. Panel c shows mean percentiles between 0.25 and 0.41 for sibling branches and for the shared past against four kinds of reference segments, all below the chance value of 0.5. Panel d shows identification rising from about 0.29 with the first meeting to 0.46 with the first eight, and from 0.33 to 0.51 with the latest eight, matching the 0.51 for all earlier meetings.}
  \label{fig:signature}
\end{figure*}

Section~\ref{sec:change} examined three established perspectives on meeting-level change and found each relevant but insufficient, while all three kept pointing back to the collaboration itself.
Finally, we ask: what, then, describes a collaboration as a unit in its own right?
We propose a collaboration's \emph{signature}, in the sense in which individuals keep persistent signatures in how they distribute their communication~\cite{saramaki2014persistence}: measures that look ordinary for any single meeting, accumulated over the meetings along a collaboration's tree, yet may together form a profile that identifies this collaboration among others.
The profile itself is deliberately simple, based on averaging the 7 composite measures of communication style and the 4 agenda measures of \S\ref{sec:change}, each rank-normalized over all meetings.
What may make it a signature is behavior: if a collaboration carries its own way of working, its profile should be distinctive beyond its subject, people, and the size and length of its meetings; part of it should be specific to the collaboration, e.g., beyond a direct pooling of its participants; it should be carried along the recovered tree into the branches the collaboration splits into.

To test collaboration-specific signatures,
we focus on the 965 collaborations with at least six meetings, allowing the possible formation of stable signatures. We build signatures from the earlier half of the meetings and ask whether each of the 6,242 later meetings lies closer to its own collaboration's signature than to those of nine competing collaborations (chance is 10\%; Appendix~\ref{app:prediction-details}).

\subsection{A signature distinguishes each collaboration}
\label{subsec:distinctive}

Against competitors drawn at random, a later meeting picks out its own collaboration 47.5\% of the time, 4.8 times chance (Figure~\ref{fig:signature}a).
Restricting competitors to the same main subject lowers this only to 44.9\%, 4.5 times chance, or competitors that also met in the same weeks, with similar headcount and duration, to 42.5\% (95\% CI 40.8 to 44.1).
Even against a collaboration that includes at least two of its participants, a meeting lies closer to its own signature in 67.2\% of comparisons, against 50\% by chance.
The signature resides in the combination of signals: interaction structure alone identifies the collaboration 33.1\% of the time, language 31.3\%, and the agenda 24.3\%, and no single measure exceeds 20.0\%.
Each kind of signal thus carries part of the signature, and a collaboration is recognized best by their configuration, by how its meetings are conducted together with what they take up.

\subsection{Part of the signature belongs to the collaboration}
\label{subsec:collaboration-specific}

Distinctiveness may still be borrowed from participants, who bring their habits to every collaboration they join.
With linked records, we can separate such composition from properties that emerge through interaction~\cite{kozlowski2000multilevel}.
We predict each collaboration's profile over its later meetings from its participants' profiles in other collaborations and from the typical profile of its main subject, correct both predictors for their own noise, and compare the result with how reliably the profile itself is measured (Appendix~\ref{app:prediction-details}).
Participants and subject together account for about two thirds of the reliable profile, mostly through participants, and the remaining 34.3\% (95\% CI 28.6 to 40.2) is specific to the collaboration (Figure~\ref{fig:signature}b).
Without any correction, the same holds out of sample: a collaboration's earlier meetings add 0.139 to the $R^2$ of its later profile beyond participants and subject, which add 0.068 beyond the earlier meetings.

The two parts differ in kind: the style of talk is mostly brought in by participants, with only 23.6\% of lexical richness and 27.7\% of first-person plural framing specific to the collaboration, whereas how a collaboration organizes its floor and its agenda is more its own, including 46.8\% of the planned share and 39.3\% of the inequality of airtime.
Against competitors that share its participants, a meeting on a new main subject lies closer to its own signature about as often as one on the usual subject (65.7\% against 68.4\%).
A collaboration thus keeps its signature through a change of subject, and loses part of it when its people change (Appendix~\ref{app:prediction-details}).

\subsection{Signatures are carried along the tree}
\label{subsec:inherited}

A signature that belongs to the collaboration should also follow the recovered structure into the branches a collaboration splits into (Section~\ref{subsec:meeting-layer}).
At 1,969 branch points, we compare the second branch with segments of the same length from other collaborations starting within two weeks of it, and record its percentile among them in distance to the first branch, 0.5 when it is no closer than they are.
Sibling branches fall at a mean percentile of 0.316, and each branch lies closer still to the meetings before the split, at 0.249 (Figure~\ref{fig:signature}c).
Both hold for communication style alone, which the linkers never see (0.312 and 0.242), for branches that share less than a third of their participants (0.355 and 0.284; 1,054 branch points), and against segments of other collaborations joined by at least two of the branch's participants within four weeks (0.412 and 0.360; 281 branch points).
A branch thus inherits the way of working of the collaboration it leaves, beyond the people who carry it over.

\subsection{Signatures emerge early, sharpen over time}
\label{subsec:horizon}

How much of a record does a signature need?
Holding out the last four meetings of the 331 collaborations with at least twelve, the first meeting alone identifies the collaboration 29.2\% of the time, the first eight meetings 46.4\%, the latest eight 50.9\%, and all earlier meetings 51.0\% (Figure~\ref{fig:signature}d).
A collaboration's first meetings thus reveal much of its signature, sharpening and only slowly drifting; correspondingly, the signature's advantage over the latest meeting in predictions widens as the record grows (Appendix~\ref{app:prediction-details}).

\subsection{The signature as a model of change}
\label{subsec:signature-model}

These properties also make the signature a parsimonious model of how a collaboration changes.
It anticipates the next meeting better than the latest meeting does on every one of the ten measures (mean AUC 0.730 against 0.686), adds more to the latest meeting and to profiles of the participants and the subject (0.035) than they add to it (0.006), and, combined with what the subject, participants, and position predict, fits whole trees better than the three perspectives combined (mean $R^2$ 0.274 against 0.240; Appendix~\ref{app:prediction-details}).

Taken together, a collaboration carries a slowly drifting signature, partly brought in by its participants and partly its own, against which it re-divides its work, passes it between subjects, and records it as completed when it ends.
The account also bears on the reconstruction: had the recovered trees joined unrelated meetings, their signatures would identify later meetings less well, carry less into branches, and sharpen less as meetings are added.

\section{Discussion}
\label{sec:discussion}

Recovering collaborations and their signatures moves the unit of observation from the single meeting or tie to the collaboration over its long horizon, with consequences for understanding collaborative work and for designing the Web systems that support it.

\subsection{Understanding collaborations as they unfold}
\label{subsec:implications-science}

\paragraph{Collaborations with signatures of their own.}
Individuals keep persistent signatures in how they distribute their communication~\cite{saramaki2014persistence}, and our results show that collaborations keep signatures too.
Multilevel theory distinguishes properties that a group composes from its members from those that emerge from their interaction~\cite{kozlowski2000multilevel}, and research on collective intelligence finds that how evenly a group shares its conversational turns predicts its performance better than its members' abilities~\cite{woolley2010evidence,riedl2021quantifying}.
The signature locates this distinction in everyday work: about two thirds of it is brought in by participants, mainly as a style of talk, whereas how a collaboration organizes its floor and its agenda is more its own, including 46.8\% of the planned share and 39.3\% of the inequality of airtime.
Conversational properties that laboratory studies treat as group-level thus persist over months in the field, and since a signature emerges within a collaboration's first meetings, those meetings are a promising point for interventions that aim to shape how it works.

\paragraph{Continuity beyond fixed members and tasks.}
Team research has increasingly questioned the picture of a bounded team with fixed members and a fixed task~\cite{mortensen2018rethinking,edmondson2012teaming}, and has long called for evidence on how groups change over time, which laboratory studies and surveys capture only in short windows~\cite{cronin2011dynamics,kozlowski2018unpacking}.
The cross-meeting network offers such evidence at the scale of an organization: most collaborations proceed as lineages of meetings that occasionally branch, along which work changes faster than people, as the subject changes at 43.4\% of links while about three quarters of participants return whether it changes or not.
A lineage's continuity thus lies less in a fixed membership or task than in a way of working, captured by its signature, which survives a change of subject and passes into branches staffed largely by other people, much like an organizational routine~\cite{feldman2003reconceptualizing}.
Linked records also make it possible to test, at scale and against outcomes, whether such lineages gain from the familiarity of a stable core~\cite{huckman2009team} and from the knowledge that newcomers bring in~\cite{choi2005old}.

\subsection{Designing for long-horizon collaboration}
\label{subsec:implications-systems}

\paragraph{Memory beyond the latest meeting.}
Less than half of a meeting's agenda continues its predecessor, an earlier meeting reveals nearly as much about the next one as the latest, and the signature, accumulated over all earlier meetings, anticipates the next meeting better than the latest does.
Memory for collaborations thus needs a longer horizon than memory for individuals, which usually favors recent items, as in agents that weight memories by recency~\cite{park2023generative} and in repeat consumption on the Web~\cite{anderson2014dynamics}.
Since a new meeting depends on the tree it joins more than on its parent, network models should likewise condition on the whole tree, which first-order models do not~\cite{rosvall2014memory,scholtes2017network}.
Assistants and retrieval pipelines that draw on earlier meetings~\cite{lewis2020retrieval,packer2023memgpt}, whose gains have so far come mainly in individual tasks~\cite{dillon2025shifting,butler2025microsoft}, and temporal graph models~\cite{huang2023temporal} could keep a tree-level summary such as the signature beside the recent context they condition on, and their predictions of a collaboration's next agenda, subjects, or participants should be evaluated against it, not only against the latest meeting~\cite{poursafaei2022towards}. More generally, AI agents and assistants that join a collaboration could adapt from its first meetings to how it organizes its floor and agenda, instead of averaging over the individuals present.

\paragraph{Tracking and governance.}
Following a piece of work means following it through the splits and merges of the agenda layer and across the branches of its collaboration, and analytics that assess one meeting at a time~\cite{hosseinkashi2024meeting} gain from placing it within its collaboration, where work is recorded as completed mainly at the end.
Tracking collaborations beyond the one organization studied here could further show whether ways of working form as early and persist as strongly in startups, other industries, and other cultures, where team size~\cite{wu2019large} and cultural norms~\cite{gelfand2007cross} shape how groups form and work.
At the same time, communication style alone identifies a meeting's collaboration at more than four times chance, so interaction metrics aggregated across meetings can identify groups, and monitoring or ranking teams by them would extend the algorithmic control of work documented in organizational research~\cite{kellogg2020algorithms}; any deployment calls for aggregation, transparency, and consent.

\section{Conclusion}
\label{sec:conclusion}

This paper set out to recover long-horizon collaboration from the discrete, unlabeled records that meeting platforms already keep.
The recovered cross-meeting network takes a dual shape: meetings form trees in which branching is more common than merging, carrying a dense directed acyclic graph of work that is re-divided at every step.
Task division, team composition, and temporal continuity each explain part of how these networks change, and all point to the collaboration itself, whose signature is distinctive beyond its subject and people, partly its own, inherited along its tree, and visible within its first meetings.
We hope this work encourages Web systems, and the research on collaborative work, to treat collaborations as units that persist, accumulate, and change over time.

\section*{Ethics and Privacy Statement}

This work involves no experiments with human participants; it analyzes existing meeting records of one organization, collected with employee consent and analyzed with institutional privacy approval within that organization's environment.
Names served only as keys for exact matching, no individual is named or scored, no record text appears, and all results are aggregates; Appendix~\ref{app:data} details these safeguards, and Section~\ref{subsec:implications-systems} notes the group-level identifiability of communication style.

\bibliographystyle{ACM-Reference-Format}
\bibliography{custom}

\appendix

\setcounter{figure}{0}
\renewcommand{\thefigure}{A\arabic{figure}}
\setcounter{table}{0}
\renewcommand{\thetable}{A\arabic{table}}

\section{Data and anonymization}
\label{app:data}

The records come from the online meeting platform of a large multinational technology company and comprise all 45,426 transcribed meetings of the groups in the study between 1 August and 31 December 2025, each with its start time, duration, roster, and a transcript of time-stamped, speaker-labeled utterances.
Rosters and speaker labels hold display names, transcripts reproduce speech verbatim, and each extracted agenda item holds a free-text description and the name of the person to whom it is attributed; we did not link the records to any other source of personal information.
Agenda extraction, subject labeling, and both link judgments use GPT-5.5 on a privacy-compliant internal deployment, and a locally run open-source sentence encoder (all-MiniLM-L6-v2) helps propose candidate item pairs.
The records were collected with employees' awareness and consent under the organization's data-use agreements, the analysis was reviewed and approved through institutional privacy review, and all processing took place in secure, access-controlled compute environments within the organization.

\paragraph{Agenda extraction and link judgments.}
The model reads each speaker-labeled transcript and returns a list of atomic agenda items following a 5W1H style, each with its content (the topic, task, decision, or action), the person it is assigned to or owned by (or the team), any time mentioned, a description of how and why, its stage (reviewing earlier work, in the meeting, or planned after it), and its status (completed, ongoing, or unknown), with stage and status judged independently.
Items are split whenever the same issue has an in-meeting and a planned aspect, members are assigned separate tasks on the same topic, or the content bundles distinct topics, and routine meeting operations are excluded.
A second pass labels each item with one of fourteen subject fields (software engineering; AI and agents; data and analytics; program and project management; product and design; infrastructure; marketing; reliability; release and deployment; security; testing; people and organization; business and finance; compliance) or as other, and assigns each meeting a main subject from the same list.
For meeting links, the model rates continuation from 1 to 5 and must cite specific threads carried over from the earlier meeting; topical similarity, recurring-meeting overlap, or shared participants alone do not count.
These rules summarize the original prompts; the full prompts will be released with the full version of the paper.

\paragraph{Identities.}
Names serve only as keys for exact matching (comparing rosters, finding participants' main subjects and measures in other collaborations, and selecting competitors that share participants), so keyed pseudonyms would leave every result unchanged.
No analysis models, scores, or compares individuals, no name, meeting title, or record text appears, and every quantity is an aggregate except the collaborations of Figures~\ref{fig:reconstruction} and~\ref{fig:style-cases} as instances.

\section{Network and component details}
\label{app:reconstruction}
\label{app:task-details}
\label{app:team-details}

\begin{figure*}[t]
  \centering
  \includegraphics[width=0.75\textwidth]{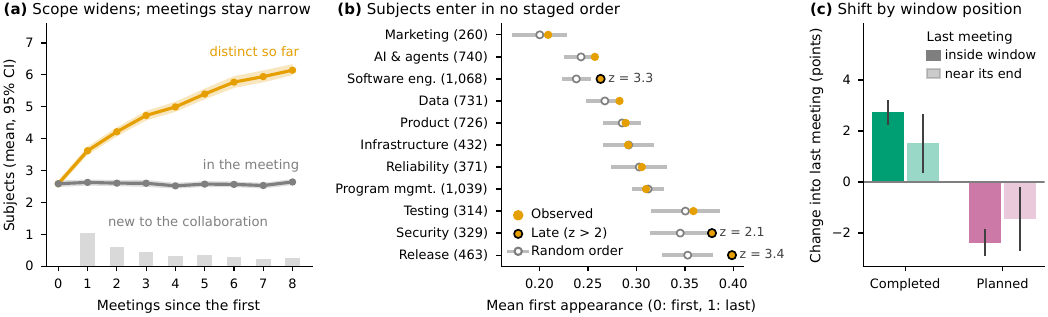}
  \caption{Task division details. Collaborations accumulate subjects while each meeting stays narrow, and subjects enter in no staged order (\S\ref{subsec:task-division}). (a) Distinct subjects so far, subjects per meeting, and new subjects (bars), for the 393 collaborations that reach a ninth meeting. (b) Mean first appearance of each subject (0: first meeting, 1: last), observed and under 200 random reorderings (gray: mean $\pm$ two standard deviations; outlined: later than that; parentheses: collaborations). (c) Change in the completed and planned shares into the last meeting, by whether it falls at least three weeks before the window closes.}
  \Description{Three panels. Panel a shows the number of distinct subjects discussed so far rising from 2.6 to 6.1 over nine meetings, while subjects per meeting stay near 2.6 and new subjects per meeting fall from about 1.0 to about 0.25. Panel b shows, for eleven subjects, the observed mean entry position next to the range expected under random reordering; observed positions lie within or near the expected ranges, with release, security, and software engineering slightly later. Panel c shows that in the last step the completed share rises and the planned share falls, by more when the last meeting ends at least three weeks before the window closes.}
  \label{fig:task-details}
\end{figure*}

\paragraph{Meeting links.}
Each meeting is attached to its most recent qualifying predecessor; lowering the acceptance threshold from 4 to 3 adds 1,394 links without changing any structural result in Section~\ref{subsec:shape}.
Scoring every other candidate predecessor that is neither an ancestor nor a descendant of the assigned one yields 8,004 additional links (3,094 within the same collaboration, 1,366 to the same recurring series in another collaboration, 2,331 to other collaborations, and 1,213 to one-off meetings), so the 19,376 tree links make up 70.8\% of all qualifying links.
A second parent is recorded as a genuine merge when it belongs to a collaboration and each parent accounts for at least three of the child's items that the other does not, judged by high-confidence item links; given the item linker's precision of 0.72, three such items would arise from independent errors alone with probability about 0.02.
This keeps 3,301 links (1,466 within a collaboration, 703 to the same recurring series, and 1,132 between collaborations).
With a threshold of one to five items, 15.5\% to 6.0\% of the meetings that continue earlier work merge, against 20.2\% to 15.1\% of the meetings with a successor that branch; counting every qualifying parent, including one-off meetings, the two shares are 22.5\% and 21.8\%.
The share of collaborations with a branch exceeds the share with a merge in every length class, by 7.2 (95\% CI 4.9 to 9.7) to 31.7 points, narrowing to 5.6 (0.0 to 10.5) among the 124 collaborations of twenty or more meetings, and merging approaches branching only ten or more steps from the root (19.8\% against 24.6\%).
Measured along the longest path through any recorded parent, depth would differ for 1.7\% of meetings and height for 0.8\% of collaborations.
Across meetings continued by two or more meetings, 16.7\% of the 28,358 items passed on are taken up by more than one branch.

\paragraph{Item links.}
For each item of a linked meeting, the candidate earlier items are the union of the 12 items of the predecessor most similar by sentence-embedding cosine and the 12 most similar by TF-IDF cosine over content and description.
The LLM judges all candidates of a meeting pair in one prompt and returns a confidence for each, and only high-confidence links are kept; against labels that the model itself produced, they have a precision of 0.72.

\paragraph{Task division.}
In the entry-order test (Figure~\ref{fig:task-details}b), each collaboration's meetings are randomly reordered 200 times; software engineering, security, and release enter significantly later than under reordering ($z$ of 3.3, 2.1, and 3.4), and no subject enters significantly earlier.
Between consecutive meetings that both have successors (11,131 steps), the completed share rises by 0.34 points per step (95\% CI 0.14 to 0.55); in the step into a collaboration's last observed meeting (8,245 steps), it rises by 2.55 points (2.10 to 2.98) and the planned share falls by 2.25 (1.79 to 2.71).
These changes are 2.75 and $-$2.40 points when the last meeting falls at least three weeks before the window closes, and 1.53 and $-$1.43 points when it falls within the last three weeks (Figure~\ref{fig:task-details}c).

\paragraph{Team composition.}
Roster carry-over cannot fall below one third, because the linker requires a roster Jaccard similarity of at least 0.33.
Among participants who join at a change of subject (6,714 cases), 32.6\% (95\% CI 31.1 to 34.1) come mainly from the incoming subject and 23.5\% (22.1 to 24.7) from the outgoing one, against 27.7\% and 29.4\% among the 18,121 participants present at both.

\paragraph{Measures of the agenda.}
The four agenda measures of Section~\ref{sec:change} are the number of items, the shares of items whose status is completed and whose stage is planned after the meeting, and the focus of the agenda, the mean pairwise cosine similarity of its items' sentence embeddings, which is high when the items address one topic and low when they range widely.

\paragraph{The floor and its measures.}
The seven measures of communication style describe the floor in the sense of conversation analysis: who holds the turn to speak, for how long, and how the turn passes from one participant to the next~\cite{sacks1974simplest}.
A transcript records the floor as a sequence of speaker-labeled utterances, which we treat as turns.
Two consecutive turns by different speakers form a \emph{handoff}, and the handoffs of a meeting form a small network in which two speakers are linked whenever one has spoken right after the other.
Each measure averages two or three standardized features that address the same construct.
Turn length describes how long a participant holds the floor, combining words and seconds of speech per turn with the share of turns of fifty or more words; long turns are typical of walkthroughs and extended explanations, short turns of quick exchanges.
Overlapping talk describes how closely participants follow one another, combining the share of turns that begin before earlier talk has ended with the share of turns of one or two words, such as brief acknowledgments.

Unequal airtime and a centralized floor describe how the floor is divided, from two sides.
Unequal airtime concerns how much each participant speaks, combining the Gini coefficients of speaking time, turns, and words across speakers.
Speaking is commonly concentrated: the median meeting has four speakers, but the median effective number of speakers, the inverse of the summed squared shares of speaking time, is 2.6.
A centralized floor concerns through whom the talk passes.
It combines the share of handoffs that involve the speaker with the most handoffs, the \emph{hub}; the share of handoffs between the busiest pair of speakers; and the Freeman degree centralization of the handoff network~\cite{freeman1978centrality}, which is 1 when the hub is linked to every other speaker and no others are linked to each other, and 0 when every speaker is linked to equally many others.
A high value describes a floor organized around one participant, as in the centralized, star-shaped communication networks of early experiments on group problem solving~\cite{bavelas1950communication}, and a low value a floor on which participants take the turn from one another directly.
Ties with the hub differ in kind from the others: across the 26,867 meetings with four or more speakers, the share of handoffs on a tie that are matched by handoffs in the opposite direction averages 0.94 for ties with the hub and 0.78 for ties between two other speakers, and it is higher for the hub's ties in 94\% of these meetings.
The hub thus tends to exchange the floor back and forth with each participant, as a coordinator who responds to every contribution would, while the other participants more often pass the turn on.
The two measures of division are related but distinct: they correlate at 0.39 across the 965 collaboration profiles of Section~\ref{sec:signature}, and 37.1\% of meetings lie on opposite sides of the median on the two.
A moderator who speaks briefly between contributions, for instance, centralizes the floor without taking much of the airtime.

\paragraph{Measures of language.}
Lexical richness combines the moving-average type-token ratio over windows of 100 words, which measures the variety of vocabulary independently of the amount of talk~\cite{covington2010cutting}, with the share of words of seven or more letters; it separates varied and technical vocabulary from plain and repetitive talk.
Hedging and politeness describe the interpersonal stance of talk, combining the rates of hedges (e.g., maybe, perhaps, probably), of politeness and appreciation words (e.g., thanks, please, sorry), and of questions; tentative and deferential exchanges score high, direct and declarative ones low~\cite{danescu2013computational}.
First-person plural framing is the share of first-person pronouns that are plural (we, us, our, as against I, me, my), and indicates whether participants speak as a group or as individuals.
The word lists are short dictionaries in dictionary-based text analysis~\cite{pennebaker2015development}.
The six measures of communication in the next-meeting tests are single features: words per turn, the number of speakers, the share of utterances beginning before earlier talk ended, first-person plural framing, the share of turns ending in a question, and the Gini coefficient of speaking time.

\begin{figure}[t]
  \centering
  \includegraphics[width=0.95\columnwidth]{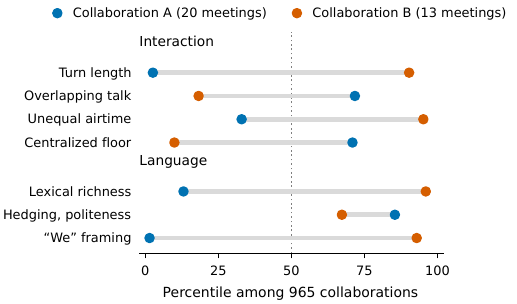}
  \caption{Case study of two collaborations on the seven measures of communication style. The two lie on opposite sides of the median on six of the seven measures: A holds short, overlapping turns routed through one participant, and B long turns with unequal airtime but no central speaker (\S\ref{subsec:team-composition}). Points are percentiles of each collaboration's profile, averaged over its meetings, among the 965 collaborations with at least six meetings; B is the collaboration with at least twelve meetings farthest from A; dotted line shows median.}
  \Description{A dot plot with seven rows, four of interaction and three of language. In each row, a blue point marks collaboration A and an orange point collaboration B, as percentiles from 0 to 100. A lies at the 3rd percentile in turn length, the 72nd in overlapping talk, the 33rd in unequal airtime, the 71st in centralized floor, the 13th in lexical richness, the 85th in hedging and politeness, and the 1st in first-person plural framing; B lies at the 90th, 18th, 95th, 10th, 96th, 67th, and 93rd.}
  \label{fig:style-cases}
\end{figure}

\paragraph{Case study of two collaborations.}
Figure~\ref{fig:style-cases} places two collaborations on the seven measures of communication style, as percentiles of their profiles, averaged over all their meetings, among the 965 collaborations of Section~\ref{sec:signature}: an example collaboration with a hub-centralized floor (A, 20 meetings) and, among the 331 collaborations with at least twelve meetings, the one whose percentiles lie farthest from it (B, 13 meetings).
A holds short, overlapping turns routed through one participant (percentiles 3, 72, and 71 for turn length, overlapping talk, and a centralized floor), in plain words and first-person singular framing (13 and 1).
B takes long turns with highly unequal airtime, yet its handoffs do not pass through any single participant (90, 95, and 10), and its language is rich and framed as ``we'' (96 and 93).
The two collaborations lie on opposite sides of the median on six of the seven measures; both lie above it in hedging and politeness (85 and 67).
An assistant that adapts to how a collaboration organizes its floor (Section~\ref{subsec:implications-systems}) would thus meet quick exchanges routed through one participant in A, and long turns from a few participants in B.

\section{Signature and prediction details}
\label{app:prediction-details}

\paragraph{Signature tests.}
The eleven measures are rank-normalized over the 45,350 meetings that have all of them, and distances between a meeting and a signature are Euclidean.
Matched competitors are the nine collaborations with the same main subject that are nearest in the median date, headcount, and duration of their earlier meetings.
For the format adjustment, each measure is regressed on the logarithms of headcount and duration, their squares, and their product, and the residuals are rank-normalized again; headcount and duration explain 52.6\% of the variance of agenda length, 31.1\% of unequal airtime, and at most 11.1\% of any other measure, and with adjusted measures, identification against matched competitors is 41.1\%.
Comparisons with collaborations that share at least two participants are pairwise (7,591 pairs in 542 collaborations); for meetings attended mostly by newcomers, the own signature is closer in 50.7\% of pairs (95\% CI 40.9 to 62.8).

\paragraph{Collaboration-specific share.}
For 956 collaborations, the participant predictor averages, over the later meetings, each participant's mean profile in other collaborations, and the subject predictor the mean profile of the main subject in other collaborations.
Each predictor's reliability is estimated by splitting all meetings at random into halves, rebuilding the predictor from each, and applying the Spearman-Brown correction (mean 0.692 for participants and 0.986 for subjects).
The profile's own reliability is the correlation between its earlier and later halves (mean 0.592); because drift lowers this correlation, the estimate errs toward a smaller collaboration-specific share.
Without the correction for predictor noise the share is 52.0\%, and with format-adjusted measures it is 35.2\% (27.6 to 40.5).

\paragraph{Branches and groupings.}
Reference segments are up to 40 runs of consecutive meetings of the second branch's length, from other collaborations and starting within 14 days of its first meeting; subject-matched segments also share its main subject (1,942 branch points; mean percentiles 0.333 for the sibling branch and 0.265 for the shared past), and people-matched segments come from other collaborations that held a meeting with at least two of the branch's participants within 28 days.
Attendance-only trees link each meeting to the most recent earlier meeting within 21 days that shares at least two attendees with a roster Jaccard similarity of at least 0.33, without the content judgment. Within them, the mean of a meeting's other meetings fits each of the ten measures less well than within the collaborations of the cross-meeting network (mean $R^2$ 0.212 against 0.232; difference 0.021, 95\% CI 0.016 to 0.024), and they anticipate the next meeting less well (mean AUC 0.702 against 0.712; difference 0.011, 95\% CI 0.008 to 0.013).

\paragraph{Next-meeting models.}
The task view has nine agenda features (e.g., the number of items, their focus, and the shares at each stage and status), the team view thirty transcript features (turn-taking, overlapping speech, concentration of speaking time, register, and the number of participants), and the trajectory view nine features (e.g., depth, intervals between meetings, and the subjects and participants seen so far).
Targets are defined at the 11,131 meetings that continue an earlier meeting and are themselves continued: a \emph{level} asks whether the next meeting lies above the corpus median, and a \emph{departure} whether it lies above the collaboration's own mean so far.
Models are regularized logistic regressions on standardized features, evaluated by mean AUC under five-fold cross-validation grouped by collaboration, with views of more than eight features reduced to eight principal components.
The comparisons of Sections~\ref{subsec:temporal-continuity} and~\ref{subsec:signature-model} use the 7,867 positions with at least three meetings so far and another meeting of the same collaboration.
At these positions, the signature reaches a mean AUC of 0.730 over the ten measures, against 0.686 for the latest meeting, 0.638 for profiles of the participants, and 0.549 for the subject in other collaborations; adding the signature to the other three raises the AUC by 0.035 (95\% CI 0.031 to 0.039), adding them to the signature by 0.006, and the task, team, and trajectory views add 0.005 to 0.021.
The signature improves on the latest meeting for every group of targets, with AUCs of 0.670 and 0.623 for agenda levels and departures and 0.757 and 0.589 for those of communication, against 0.634, 0.562, 0.721, and 0.542.
Its advantage for levels widens from 0.021 (95\% CI 0.017 to 0.025) with two meetings known to 0.053 (0.043 to 0.060) with more than ten.
For new subjects and participants, the collaboration's composition so far (subjects and participants seen, meetings held) reaches AUCs of 0.679 and 0.719 (0.690 and 0.713 under random reordering), against 0.548 and 0.610 for whether the latest meeting brought one.
\paragraph{Tree-level fit and agenda overlap.}
The model of Section~\ref{subsec:signature-model} reaches a mean $R^2$ of 0.274, against 0.240 for the three perspectives combined (difference 0.035, 95\% CI 0.029 to 0.040) and at most 0.188 for any one of their predictors, on all ten measures.
The fit covers 16,249 meetings in the 2,979 collaborations with at least three meetings; the subject, participant, and depth predictors average meetings outside the meeting's own collaboration, the adjacent predictor averages the meeting it continues and those that continue it, and models are linear regressions evaluated by out-of-fold $R^2$ under five-fold cross-validation grouped by collaboration.
The signature alone reaches 0.232, and it and the perspective predictors each add to the other (0.028 and 0.035).
The full model fits, within each training fold, a linear prior from subject, participants, and depth, adds the mean residual of the collaboration's other meetings, shrunk toward the prior by the reliability of that mean, and then adds the adjacent meetings; after the prior, 22.1\% of the remaining variance lies between collaborations.
Across 19,374 meeting links, the shares of a meeting's items that continue its predecessor (46.3\% on average) and of the predecessor's items taken up (27.5\%) range from 43.0\% to 49.1\% and from 24.0\% to 29.5\% as the gap between meetings grows from under a day to three weeks (rank correlations with the gap $-$0.048 and $-$0.055).

\end{document}